\documentclass[10pt,conference]{IEEEtran}
\usepackage[T1]{fontenc}
\usepackage{cite}
\usepackage{amsmath,amssymb}
\usepackage{graphicx}
\usepackage{booktabs}
\usepackage{url}
\usepackage[hidelinks]{hyperref}
\hypersetup{pdftitle={Procrustes-Bering-Sheaft: Automated Model Discovery and Virtual Crash Tests for Microservices},pdfauthor={Anatoly A. Krasnovsky}}
\newcommand{\toolchain}{Procrustes--Bering--Sheaft}
\begin{document}
\title{Procrustes--Bering--Sheaft: Automated Model\\Discovery and Virtual Crash Tests for Microservices}
\author{\IEEEauthorblockN{Anatoly A. Krasnovsky}
\IEEEauthorblockA{Innopolis University and MB3R Lab\\
Innopolis, Russia\\
a.a.krasnovsky@gmail.com\\
ORCID: 0000-0001-6842-7340}}
\maketitle

\begin{abstract}
Procrustes--Bering--Sheaft automates lightweight model discovery and virtual crash tests for microservices. Each test estimates operation availability by Monte Carlo sampling of service and communication failures. Procrustes checks deployment and tracing evidence; Bering discovers dependencies and incorporates replica declarations; Sheaft evaluates declared operation-success conditions. Independent command-line interfaces exchange reusable, versioned artifacts. The toolchain extends an earlier graph-based approach with mandatory-call contracts, failure-tolerance sweeps, baseline comparison, component-importance analysis, and CI/CD gates. We demonstrate the workflow on one HTTP operation. Six controlled execution states exposed two mismatches between reachability and observed success; an explicit required-call predicate matched all six. Two modelled one-replica additions produced different decisions: recommendation retained an availability-target violation, whereas catalog restored the baseline estimate of 0.979900 and grid tolerance of 0.12. Analytical oracles and controlled tests verify the implementation; archived replay checks numerical continuity with two earlier studies.

%Tool and instructions: \url{https://github.com/MB3R-Lab/mb3r-software-complex-repro}.

%Video: \url{https://youtu.be/V8y0aQ4l7PM}.
\end{abstract}

\section{Introduction}
Virtual crash tests assess operation availability under service and communication failures on a reconstructed model. Our lightweight representation comprises dependencies, replicas, operation-success predicates, and a distribution over component states. Platform engineers compare configurations and apply the same checks in continuous integration and delivery (CI/CD); researchers repeat model-based analyses.

Earlier studies evaluated graph discovery and replica-aware simulation on DeathStarBench Social Network and the OpenTelemetry Demo~\cite{krasnovsky2026nier,krasnovsky2026aina}. A companion study develops identification algorithms and formal analysis for lightweight availability models~\cite{formal}. This paper presents \toolchain{}, a three-tool implementation of the discovery-and-simulation approach with expanded analytical capabilities~\cite{artifact}.

The extensions cover discovery-precondition diagnostics, mandatory-call contracts, failure-tolerance sweeps against service-level objectives (SLOs), compatible-baseline regression checks, and component-importance analysis. Versioned handoffs preserve the inputs behind each result. The demonstration connects model construction, execution checks, and simulation in a comparison of two replica changes. Analytical oracles and archived replay verify calculations and numerical continuity.

\section{Automated Discovery and Failure Simulation}
\begin{figure*}[t]
\centering
\includegraphics[width=\textwidth]{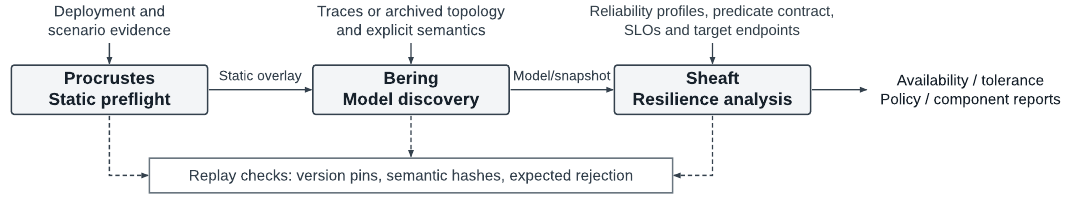}
\caption{Analysis and verification paths. Solid arrows carry stage inputs and versioned artifacts; dashed arrows connect retained outputs to replay checks. Required-call declarations enter Sheaft through an external predicate contract alongside the analysis configuration.}
\label{fig:workflow}
\end{figure*}

The stages in Fig.~\ref{fig:workflow} expose public command-line interfaces and retain their inputs and outputs~\cite{artifact}.

\subsection{Evidence Preflight and Model Discovery}
Procrustes checks deployment manifests and tracing configuration against an operation goal and Tracetest scenario. It resolves service identities and replicas, checks instrumentation and context propagation, and records evidence provenance. Unknown or conflicting evidence remains explicit. Blocked assessments retain diagnostics; \texttt{-{}-fail-on-blocked} returns nonzero. A collection plan specifies trace attributes; an overlay transfers static declarations to Bering.

Bering discovers operation-aware dependencies from traces or explicit topology evidence. It emits a model, snapshot, and optional quality reports; overlays supply deployment and endpoint declarations. Stable identifiers connect model elements, selectors, and report rows. Model/snapshot contracts use schema~1.3.0; Sheaft rejects unsupported inputs. A changed failure profile reuses the model; Bering applies revised replica declarations when rebuilding it~\cite{artifact}.

\subsection{Endpoint Success and Mandatory Calls}
Endpoint predicates specify required services and dependency modes. Immediate-response conditions exclude nonblocking asynchronous work; eventual-completion conditions can require it. An \texttt{edge\_aware} predicate checks reachability of mandatory targets over live edges. External predicate contract~1.1 additionally supports \texttt{required\_edges}: every listed call must succeed, its endpoints must be live, and its caller must be reachable. An alternative path cannot replace an explicitly required call. Sheaft validates exact Bering edge identifiers and selected modes before analysis; contract~1.0 remains supported~\cite{artifact}.

Users declare mandatory calls; the runner binds them to discovered edge identifiers, retaining every edge. Availability, sweeps, and importance share this predicate. Each channel has one connectivity state per trial, shared by repeated RPCs; execution order and payload conditions belong to the declared scenario.

\subsection{Failure Simulation and Decision Rules}
A profile fixes the failure population, distribution, trials, and seed. Sheaft estimates availability by Monte Carlo sampling of states and evaluation of the operation predicate~\cite{artifact}. With independent replica failure probability $f$, a service with $r$ replicas survives with probability $1-f^r$. Other profiles fail an exact number of eligible services or replica slots; explicit exclusions remove components from baseline sampling.

A sweep evaluates an ordered grid against an endpoint SLO and reports nominal pointwise Wilson intervals~\cite{wilson}. The \emph{certified grid tolerance} $b_L$ is the largest tested level in a contiguous prefix whose lower bounds meet the SLO. An interval spanning the SLO is indeterminate. Coupled trials preserve nested fail-stop samples; an unresolved nonmonotonic sweep cannot certify a blocking boundary.

Ordinary endpoint thresholds compare point estimates. Separate boundary rules check a minimum $b_L$ and a permitted decrease from a baseline. Sweep fingerprints check compatible axes, SLOs, confidence levels, budgets, and seed semantics. Incompatibility yields an indeterminate comparison. The regression rule compares two certified grid metrics; it is not a confidence bound on the change of a continuous threshold. Reports retain the violated rule and endpoint~\cite{artifact}.

\emph{CI/CD integration.} Sheaft returns~2 for a policy failure in \texttt{mode=fail},~1 for input, contract, configuration, or runtime errors, and~0 for pass/warn/report. A delivery pipeline requiring successful gate completion before promotion blocks candidates with nonzero exits. Sheaft's \href{https://github.com/MB3R-Lab/Sheaft/blob/9248f50ccb1bc93dc26f1ca234b4f88b52880031/docs/ci-gate.md}{\emph{CI Gate}} guide documents fail-closed handling of indeterminate boundaries and templates for GitHub Actions, GitLab CI, and Jenkins. Reusable adapters are distributed in \href{https://github.com/MB3R-Lab/mb3r-stack/releases/tag/v1.2.0}{\texttt{mb3r-stack}~v1.2.0}.

\subsection{Component Inspection}
Paired Birnbaum analysis~\cite{birnbaum} forces one service or edge live and failed while preserving other sampled states. It estimates $I_{e,i}=A_e^{i\uparrow}-A_e^{i\downarrow}$ and individual Wilson intervals from pivotal trials. With baseline component liveness $q_i$, reports also give forced-down impact $q_iI_{e,i}$ and perfect-component gain $(1-q_i)I_{e,i}$. A replicated service uses aggregate liveness $q_i$. This release supports importance for monotone predicates under independent service or replica sampling without a fault overlay. Importance informs component inspection; each proposed replica change receives a separate evaluation~\cite{artifact}.

\section{Demonstration: One Operation, Two Candidates}
Release~v1.3.3 packages Procrustes~0.1.0, Bering~1.0.0, and Sheaft~1.3.1 with fixed source revisions~\cite{artifact}. After obtaining the repository, a user needs Docker with Compose and runs
\begin{quote}\ttfamily
 docker compose run -{}-rm demo
\end{quote}
The command invokes the public tools on recorded observations and fixed candidates, verifies the reference summary, and saves results in \texttt{output/operation-walkthrough/}. Live evidence collection has a separate workflow; replay needs no application deployment.

\subsection{From an Observed Request to Its Predicate}
The scenario uses application revision \texttt{be466eb} of the OpenTelemetry Demo MB3R Stack~\cite{otel,apprevision} and the request \texttt{GET} \path{/api/recommendations?currencyCode=USD}. Catalog seed data are fixed, no product IDs are excluded, and \texttt{recommendationCacheFailure} is disabled. Success requires HTTP~200 and a nonempty product array with nonempty identifiers. PostgreSQL, flagd, frontend, and telemetry infrastructure are held healthy. The frontend calls recommendation, which obtains catalog IDs through \texttt{ListProducts}; frontend then retrieves up to four products through \texttt{GetProduct}. Both catalog calls belong to this scenario.

We collected two requests for each of six predeclared states (Table~\ref{tab:semantics}), with recovery checks between conditions. Service faults stopped the actual container. Directed-channel faults closed existing connections and prevented reconnects for the whole request. Transport timeouts would make collection inconclusive. The retained evidence contains HTTP bodies, raw Jaeger traces, deployment and image provenance, and 42 normalized spans from the two healthy requests~\cite{artifact}.

\begin{table}[t]
\caption{Two responses per fixed execution state. F: frontend; R: recommendation; C: catalog. S/F: predicted success/failure.}
\label{tab:semantics}
\centering
\footnotesize
\begin{tabular}{@{}lccc@{}}
\toprule
State & HTTP codes & Reachability & Required calls\\
\midrule
Healthy & 200, 200 & S & S\\
R stopped & 500, 500 & F & F\\
C stopped & 500, 500 & F & F\\
F $\to$ R disabled & 500, 500 & F & F\\
R $\to$ C disabled & 500, 500 & \textbf{S} & F\\
F $\to$ C disabled & 500, 500 & \textbf{S} & F\\
\bottomrule
\end{tabular}
\end{table}

Procrustes checks a goal for this same operation. Bering constructs the graph from the healthy traces and the preflight overlay. The runner verifies success labels against responses and links the responses and normalized healthy spans to their raw traces. It then binds the three declared call pairs to their discovered identifiers. On the same graph and mandatory targets, reachability incorrectly predicted success for both individual catalog-channel outages. Adding required calls matched all six observed states.

\subsection{Reviewing the Replica Changes}
The fixed protocol uses independent failures of recommendation and catalog replicas with $f=0.05$, independent channel liveness $0.995$, SLO~$0.95$, $N=40{,}000$, and seed~2719. The same settings govern importance and all candidate evaluations. The sweep grid is $\{0,0.01,0.02,0.03,0.04,0.05,\allowbreak0.06,0.08,0.10,0.12,\allowbreak0.14,0.16,0.18,0.20,0.25\}$. Rules require $b_L\geq0.10$ and, for candidates, a decrease from the baseline of at most~$0.02$~\cite{artifact}.

The observed deployment had one replica per service. We compare declared model variants: baseline $2/2$ (recommendation/catalog), reduced $2/1$, and two one-replica additions, $3/1$ and $2/2$. Both candidates have the same replica-count budget; resource costs are unspecified.

\begin{table}[t]
\caption{Model variants at $f=0.05$; replicas: R/C. Pointwise 95\% Wilson intervals; $b_L$: certified grid tolerance.}
\label{tab:candidates}
\centering
\footnotesize
\begin{tabular}{@{}lcrccc@{}}
\toprule
Model & Repl. & $\widehat A_e$ & 95\% interval & $b_L$ & Gate\\
\midrule
Baseline & 2/2 & 0.979900 & [0.9785, 0.9812] & 0.12 & pass\\
Reduced & 2/1 & 0.931200 & [0.9287, 0.9336] & 0.03 & fail\\
Add R & 3/1 & 0.934675 & [0.9322, 0.9371] & 0.03 & fail\\
Add C & 2/2 & 0.979900 & [0.9785, 0.9812] & 0.12 & pass\\
\bottomrule
\end{tabular}
\end{table}

The reduced model failed the endpoint threshold and boundary rules (Table~\ref{tab:candidates}). Birnbaum estimates were $0.934925$ for recommendation and $0.983175$ for catalog, with intervals $[0.932466,0.937301]$ and $[0.981867,0.984390]$. Their perfect-component gains were $0.002337$ and $0.049159$. Separate evaluations gave $0.934675$ for Add~R, retaining the violation, and $0.979900$, $b_L=0.12$ for Add~C~\cite{artifact}. Add~C reproduces the baseline's replica counts and sampling settings, hence the identical estimates.

\noindent\begin{minipage}[t]{\columnwidth}
For Add~R, the user can trace the failed decision to its baseline-comparison row:
\begin{quote}\footnotesize\ttfamily
"endpoint\_id":\\
\hspace*{1em}"frontend:GET /api/recommendations",\\
"certified\_tolerance": 0.03,\\
"reference\_tolerance": 0.12,\\
"regression": -0.09,\\
"max\_regression": 0.02, "status": "fail"
\end{quote}
\end{minipage}

Separate minimum and regression rules retain both diagnoses. Rejected candidates make \texttt{sheaft run} exit~2; \texttt{demo} verifies these decisions and codes, not application promotion. Users can change replica declarations or analysis settings and invoke Bering and Sheaft directly using Bering's \href{https://github.com/MB3R-Lab/Bering/blob/86db92066a5284d52b87a1000af0a0ea7e925a91/README.md\#first-run}{\emph{First Run}} and Sheaft's \href{https://github.com/MB3R-Lab/Sheaft/blob/9248f50ccb1bc93dc26f1ca234b4f88b52880031/README.md\#quickstart}{\emph{Quickstart}} and \href{https://github.com/MB3R-Lab/Sheaft/blob/9248f50ccb1bc93dc26f1ca234b4f88b52880031/docs/configuration.md}{\emph{Configuration and Schemas}} guides. The demo replays prepared candidates; users select changes.

\section{Validation and Reproducibility}
\subsection{Analytical and Controlled Checks}
The declared operation has an independent series-system oracle:
\begin{equation}
A(f,r_R,r_C)=0.995^3(1-f^{r_R})(1-f^{r_C}).
\label{eq:oracle}
\end{equation}
Channels retain one state across repeated RPCs. Exact availabilities at $f=0.05$ are $0.9801556573$ ($2/2$), $0.9334815784$ ($2/1$), and $0.9357041536$ ($3/1$). Maximum absolute discrepancy was $0.00228158$, below the fixed tolerance $0.006$; all exact values lie within the reported intervals. Even a perfectly live recommendation service leaves $0.995^3\!\cdot0.95=0.935821$ with one catalog replica. Thus Add~R cannot meet the SLO.

The validation matrix passed 15 checks: four Procrustes diagnoses, three exact Bering service/edge/endpoint-set checks, four numerical oracles, three policy outcomes, and one unsupported handoff. At $f=0.10$, 50,000 trials, and seed~2711, immediate/eventual oracle values $0.9/0.81$ yielded $0.89900/0.80914$. Policy tests distinguish pass, fail, and indeterminate reasons. Additional Sheaft tests enumerate all eight channel states of a two-caller triangle and compare nontrivial importance with a series oracle~\cite{artifact}.

\subsection{Continuity and Re-execution}
Archived replay covers Social Network's 12-service graph with and without replication and OpenTelemetry's 16-service graph with asynchronous and all-blocking interpretations~\cite{krasnovsky2026nier,krasnovsky2026aina,deathstar}. Five profiles per variant give 20 comparisons with published model values. Social Network uses fixed-count replica-slot failures; OpenTelemetry fails 2, 4, 8, 10, or 14 of 15 eligible services. Maximum absolute difference is $0.001953$. This checks numerical continuity; OpenTelemetry uses a representative graph from a study aggregating graph runs~\cite{artifact}.

Replay regenerates handoffs and checks a deterministic operation summary. Archived repeats normalize documented environmental fields, including some local provenance identifiers, and verify 16 files per variant. Source builds, container checks, and anonymous image download/execution passed for artifact revision \texttt{9971a0a}. The four-configuration replay took about 342\,s in that Linux/AMD64 CI job. The \texttt{cost} command records archived-case full-pass and saved-model time and maximum child-process RSS on Linux~\cite{artifact}.

\section{Related Work}
Palladio models component behavior, usage, and deployment~\cite{palladio}; PMX extracts PCM models from traces and metrics and integrates prediction into CI/CD~\cite{pmx}. The companion identification study~\cite{formal} reported observed mean absolute forecast errors within 0.082 percentage points of PMX on complete common campaigns. Its measured telemetry-to-first-forecast workflow was 31.73--100.24 times faster, including startup and prescribed PMX launcher waits.

MiSim's discrete-event engine models request processing, CPU scheduling, resilience mechanisms, and time-varying workloads~\cite{misim}. Th\"uring et al. extract architectures, resource demands, and resilience-mechanism information from traces for MiSim~\cite{trace_misim}. Our toolchain evaluates operation success directly on service and channel states, using discovered dependencies, replica declarations, explicit predicates, and failure profiles. This specialized representation supports model-based configuration comparison.

For architecture recovery, $\mu$TOM mines $\mu$TOSCA descriptions from deployment and interaction evidence~\cite{mutom}; GMTA groups trace paths into business flows~\cite{gmta}. Bering carries operation identities into quantitative failure analysis.

PRISM models quantify Retry and Circuit Breaker behavior~\cite{prism}; ResilienceBench compares pattern implementations under controlled workloads and faults~\cite{resiliencebench}. MicroRes profiles degradation propagation~\cite{microres}; Filibuster adds fault injection to functional tests~\cite{filibuster}. Our primary analysis executes on the dependency model.

\section{Scope and Conclusion}
Execution checks concern one fixed USD request, dataset, cache setting, and six states. Replica variants use declared probabilities; live failover and probability calibration lie outside this experiment. The model covers finite fail-stop states and excludes saturation, queue feedback, and temporal recovery. Wilson intervals quantify pointwise sampling uncertainty under the model; the combined policy has no asserted simultaneous 95\% guarantee.

\toolchain{} automates lightweight model discovery and virtual crash tests. Independent stages and versioned contracts connect evidence diagnostics to failure simulation, component inspection, configuration comparison, and CI/CD gates. The demonstration links observed dependencies and a checked success condition to two separately evaluated replica changes, with analytical oracles and archived replay verifying calculation and continuity.

\section*{Data Availability}
Release~v1.3.3~\cite{artifact} provides the pinned tools, plan, HTTP responses, raw/normalized traces, configurations, and reference reports. Docker replay and live collection have separate workflows. The release records the container digest and component commits.

\bibliographystyle{IEEEtran}
\bibliography{references}
\end{document}